\documentclass[a4paper,12pt]{article}
\usepackage{amsmath}
\usepackage{amsfonts,amssymb}
\usepackage{preprint}
\begin{document}
\numberwithin{equation}{section}
\baselineskip=14pt  
\abovedisplayskip=10pt
\belowdisplayskip=10pt
\jot=5pt
\rm
%
\def\calO{{\cal O}}
\def\calH{{\cal H}}
\def\calV{{\cal V}}
\def\calFP{{\cal FP}}
\def\boldN{{\mathbf{N}}}
\def\boldR{{\mathbf{R}}}
\def\boldC{{\mathbf{C}}}
\def\bolda{\hbox{\boldmath$\mathit{a}$}}
\def\boldc{\hbox{\boldmath$\mathit{c}$}}
\def\boldbarc{\hbox{\boldmath$\bar{\mathit{c}}$}}
\def\rvac{|\,0\,\rangle}
\def\barc{{\bar c}}
\def\barC{{\bar C}}
%
\hrule height0pt depth0pt
\vspace*{-12pt}
\rightline{\bf RIMS-1333}
\vspace*{50pt}
\centerline{\Huge 
Pseudo Cuntz Algebra and}
\vskip15pt
\centerline{\Huge 
Recursive FP Ghost System in String Theory
}
\vskip100pt
\renewcommand{\thefootnote}{\alph{footnote})}
\centerline{\large
 Mitsuo Abe\footnote{E-mail: \abemail}
 and Katsunori Kawamura\footnote{E-mail: \kkmail}
}
\vskip10pt
\centerline{\it Research Institute for Mathematical Sciences,
Kyoto University, Kyoto 606-8502, Japan}
\vskip25pt
\vskip120pt
\centerline{\itshape\bfseries Abstract}
\vskip5pt
Representation of the algebra of FP (anti)ghosts in string theory is studied
by generalizing the recursive fermion system in the Cuntz algebra constructed
previously.
For that purpose, the pseudo Cuntz algebra, which is a $\ast$-algebra 
generalizing the Cuntz algebra and acting on indefinite-metric vector spaces, 
is introduced. The algebra of FP (anti)ghosts in string theory is embedded 
into the pseudo Cuntz algebra recursively in two different ways. 
Restricting a certain permutation representation of the pseudo Cuntz algebra, 
representations of these two recursive FP ghost systems are obtained. 
With respect to the zero-mode operators of FP (anti)ghosts, it is shown 
that one corresponds to the four-dimensional representation found recently by 
one of the present authors (M.A.) and Nakanishi, while the other corresponds 
to the two-dimensional one by Kato and Ogawa.
\vfill\eject
%
%
%
\section{Introduction}
In our previous paper,\cite{AK1} we have introduced the recursive 
fermion system (RFS) which gives embeddings of the fermion algebra (or CAR) 
into the Cuntz algebra $\calO_2$ or $\calO_{2^p}$ $(p\geqq2)$. We have shown 
how the representations of the fermion algebra are obtained by restricting 
those of the Cuntz algebra. According to embeddings, we can obtain unitarily 
inequivalent representations of the fermion algebra.\cite{AK2}
In that framework, however, we can not treat unphysical fermions such as 
FP (anti)ghosts, which are defined only on the basis of the indefinite-metric 
vector space, since the Cuntz algebra is represented on Hilbert space with the 
conventional positive-definite inner product. 
Since FP (anti)ghosts play quite an important role in gauge theories and 
quantum gravity, it is very desirable for us to have a similar formulation 
to manage them.
In order to treat such unphysical fermion algebras in the same way we have 
done in the Cuntz algebra, we need to generalize the Cuntz algebra itself 
so that it acts on the indefinite-metric vector space. 
\par
As for the FP (anti)ghost fields in string theory,  their mode-decomposed 
operators satisfy the anticommutation relations with the special structure 
owing to the Hermiticity of the FP (anti)ghost fields as follows:\footnote{%
Throughout this paper, we use $^*$ to denote the Hermitian conjugate 
instead of $^\dagger$ in conformity with the notation of $\ast$-algebra. 
The minus signs appearing in rhs of \eqref{c0-barc0} and \eqref{cm-barcn} are
just for our convention. If one prefers to the plus sign, one only has to 
redefine $-\barc_n$ as $\barc_n$ for $n\geqq0$.}
\begin{eqnarray}
&& \{ c_0, \, \barc_0 \} = -I, \quad 
    c_0^* = c_0, \ \ \barc_0^*=\barc_0,              \label{c0-barc0}\\ 
&& \{ c_m, \, \barc_n^* \} 
 = \{ c_m^*, \, \barc_n \} =-\delta_{m,n}I, 
    \quad m, \, n= 1,\,2\,\ldots,                    \label{cm-barcn}
\end{eqnarray}
and other anticommutation relations vanish.
Diagonalizing \eqref{c0-barc0} and \eqref{cm-barcn}, we rewrite them into 
the following
\begin{eqnarray}
&& b_1 \equiv c_0 + \barc_0=b_1^*, \quad 
   b_2 \equiv c_0 - \barc_0=b_2^*, \\
&& \{ b_i, \, b_j \} = 2\eta_{i,j}I, \quad 
   \eta_{i.j}=\mbox{diag}(-1, \, +1), \ \ i,\,j=1,\,2, \\
&& a_{2n+1}  \equiv \frac{1}{\sqrt2}(c_n + \barc_n), \quad 
   a_{2(n+1)}\equiv \frac{1}{\sqrt2}(c_n - \barc_n), \ \ 
   n=1,\,2,\,\ldots,\\
&& \{ a_m, \, a_n^* \} = (-1)^m \delta_{m,n}I, \quad  
   m,\,n=3,\,4,\,5,\,\ldots,
\end{eqnarray}
and others vanish. Thus, the zero-mode operators satisfy the anticommutation 
relations isomorphic with the $(1+1)$-dimensional Clifford algebra.  
Therefore, we can not adopt the conventional Fock representation with 
regard to them. To overcome this difficulty, Kato and Ogawa\cite{KO} 
introduced the two-dimensional representation with respect to the zero-mode
operators:\footnote{These equations are different from the original ones 
in Ref.\,\citen{KO}) by an extra minus sign owing to our convention of 
the anticommutation relation \eqref{c0-barc0}.}  
\begin{equation}
  \begin{array}{l}
  c_0\,| \, - \, \rangle = |\,+\,\rangle, \quad
  \barc_0\,| \, + \, \rangle = - |\,-\,\rangle, \quad
  c_0\, | \, + \, \rangle = \barc_0\,| \, - \, \rangle = 0, \\[2pt]
  \langle \,\pm\, | \,\mp\,\rangle=1, \quad
  \langle \,\pm\, | \,\pm\,\rangle=0.
  \end{array}
\end{equation}
Here, $|\,\pm\,\rangle$ is an eigenvector of the FP ghost number charge 
$i Q_c$ with eigenvalue $\pm \frac{1}{2}$, thus the FP ghost numbers are 
half-integers in Kato-Ogawa theory. Since this two-dimensional representation 
has the indefinite inner product with the off-diagonal metric structure, 
their vacuum with respect to the FP (anti)ghost operators is orthogonal with 
itself. Therefore, they also introduced the ad hoc metric operator in order 
to recover the vacuum expectation values in the conventional sense. 
However, it is not admissible to introduce such a metric operator since in 
general it violates the operator Hermitian conjugation at the representation 
level. More suitable representation for the zero-mode operators is the 
four-dimensional one which is recently obtained in the exact solution to the 
operator formalism of the conformal-gauge bosonic string theory.\cite{AN}  
The four vacuum vectors are denoted by 
\begin{equation}
  \begin{array}{l}
  \rvac, \quad c_0\,\rvac, \quad 
  \barc_0\,\rvac, \quad [\barc_0,\,c_0]\,\rvac, \\[2pt]
  \langle\,0\,|\,0\,\rangle=1.
  \end{array}
\end{equation}
In this formulation, we have the genuine vacuum (cyclic vector) $\rvac$ 
with the positive norm for the FP (anti)ghost operators, hence we do not 
need to introduce a metric operator as above.
Since $\rvac$ is {\it not\/} an eigenvector of $i Q_c$, it is necessary to 
project out the zero-mode operators. Then, we have integer FP ghost numbers.
\par
The purpose of this paper is to introduce the pseudo Cuntz algebra which 
acts on the indefinite-metric vector space and study the representation 
of the FP ghost algebra in string theory by restricting that of the pseudo 
Cuntz algebra. We construct two embeddings $\varPhi_1$ and $\varPhi_2$ of 
the FP ghost algebra $\calFP$ into the pseudo Cuntz algebra $\calO_{2,2}$ by 
extending our previous results for the recursive fermion system:
\begin{equation}
 \varPhi_i : \calFP \hookrightarrow \calO_{2,2} \quad i=1,\,2.
\end{equation}
We show that one of them has a four-dimensional representation with respect 
to the zero-mode operators and the other a two-dimensional one, when a 
certain representation of the pseudo Cuntz algebra is restricted.
These representations are unitarily inequivalent with each other.
\par
The present paper is organized as follows. 
In Sec.\ 2, our previous results for the recursive fermion system in the Cuntz 
algebra are reviewed. In Sec.\ 3, the pseudo Cuntz algebra is introduced.
In Sec.\ 4,  the recursive construction of embeddings of the FP ghost algebra
in string theory into the pseudo Cuntz algebra is presented.
In Sec.\ 5, the representation of our FP ghost system is constructed.
The final section is devoted to discussion. 
In Appendix, some embeddings among pseudo Cuntz algebras are presented 
in brief.
\vskip20pt
%
%
%
\section{Recursive Fermion System in Cuntz Algebra}
\subsection{Definition}
We show a method of systematic construction of an embedding of the fermion
algebra into the Cuntz algebra $\calO_{2^p}$.
\par 
The Cuntz algebra\cite{Cuntz} $\calO_d$ is a $C^*$-algebra generated by 
$s_i$ $(i=1,\,2,\,\ldots,\,d)$ satisfying the following relations
\nopagebreak
\begin{eqnarray}
\noalign{\vskip-2pt}
&& s_i^*\, s_j = \delta_{i,j}\,I,                    \label{CR1}\\[-2pt]
&& \sum_{i=1}^d s_i \, s_i^* = I,                    \label{CR2} 
\end{eqnarray}
\eject\vfill\noindent
where $I$ is the unit (or the identity operator). 
We often use the brief description such as 
$s_{i_1\cdots i_m}\equiv s_{i_1}\cdots s_{i_m}$,
$s_{i_1\cdots i_m}^* \equiv s_{i_m}^*\cdots s_{i_1}^*$ and 
$s_{i_1\cdots i_m;\, j_n\cdots j_1}\equiv 
s_{i_1}\cdots s_{i_m} s_{j_n}^*\cdots s_{j_1}^*$.
\par
The fermion algebra (or CAR) is a $C^*$-algebra generated by 
$a_n\ (n=1,\,2,\,\ldots)$ satisfying
\begin{eqnarray}
\noalign{\vskip-8pt} 
&& \{ a_m, \, a_n \} = 0, \quad 
   \{ a_m, \, a_n^* \} = \delta_{m,n}I, 
   \quad m,\,n=1,\,2,\,\ldots.                       \label{CAR}
\end{eqnarray}
\par
We construct embeddings of CAR into $\calO_{2^p}$ as a 
$\ast$-subalgebra.\cite{AK1} For this purpose, we introduce 
$\bolda_1,\,\bolda_2,\,\ldots,\,\bolda_p\in \calO_{2^p}$,
a linear mapping $\zeta_p:\calO_{2^p}\to\calO_{2^p}$, and a unital 
(i.e., preserving $I$) endomorphism $\varphi_p$ on $\calO_{2^p}$. A set 
$R_p=(\bolda_1,\,\bolda_2,\,\ldots,\,\bolda_p;\, \zeta_p,\,\varphi_p)$ 
is called a {\it recursive fermion system of order $p$\/} (RFS$_p$) in 
$\calO_{2^p}$, if it satisfies 
\begin{alignat}{3}
&\mbox{i) seed condition:} &\ \ &\{\bolda_i, \, \bolda_j\}=0, \quad 
   \{\bolda_i,\,\bolda_j^*\}=\delta_{i,j}I,   
   \label{rfsp-1}\\
&\mbox{ii) recursive condition:} &\ \ &\{\bolda_i,\,\zeta_p(X)\} = 0, \quad
  \zeta_p(X)^*=\zeta_p(X^*),  \quad 
  \ X \in \calO_{2^p},      \label{rfsp-2}\\
&\mbox{iii) normalization condition:} &\ \ &
   \zeta_p(X)\zeta_p(Y)=\varphi_p(XY), \quad 
    X, \,Y \in \calO_{2^p},\label{rfsp-3}
\end{alignat}
with none of $\bolda_i$ been expressed as $\zeta_p(X)$ with 
$X\in\calO_{2^p}$. An embedding $\varPhi_{R_p}$ of CAR into $\calO_{2^p}$ 
associated with $R_p$ is defined by
\begin{eqnarray}
&&\begin{array}{c}
\varPhi_{R_p} : \mbox{CAR} \hookrightarrow \calO_{2^p},    \\[5pt]
\varPhi_{R_p}(a_{p(n-1)+i}) \equiv \zeta_p^{n-1}(\bolda_i), \quad 
   i=1,\,\ldots,\,p,\, n=1,\,2,\,\ldots. 
\end{array}                                          \label{rfsp-4}
\end{eqnarray}
It is, indeed, straightforward to show that \eqref{rfsp-4} satisfies 
the anticommutation relations isomorphic with \eqref{CAR}.
\par
In the following, we give examples called the {\it standard \/} RFS$_p$ 
for the cases $p=1, \, 2$.
\begin{list}{}{\topsep=3pt\itemsep=0pt\parsep=0pt}
\item[(1)] $p=1$,  the standard RFS$_1$ in $\calO_2$, 
$SR_1=(\bolda; \, \zeta,\,\varphi)$:
\begin{eqnarray}
&& \bolda \equiv s_1\,s_2^*,                         \label{uhf-rfs-1}\\
&& \zeta (X) \equiv s_1 X s_1^* - s_2 X s_2^*,       \label{uhf-rfs-2}\\
&& \varphi(X) \equiv \rho(X) 
              \equiv s_1 X s_1^*+s_2 X s_2^*,        \label{uhf-rfs-3}
\end{eqnarray}
where $\rho$ is the canonical endomorphism of $\calO_{2}$.
The embedding $\varPhi_{SR_1}$ of CAR into $\calO_2$ associated with $SR_1$ 
is given by
\begin{equation}
\varPhi_{SR_1}(a_n)=\zeta^{n-1}(\bolda), \quad 
   n=1,\,2,\,\ldots.                                 \label{uhf-rfs-4}
\end{equation}
\item[(2)] $p=2$, the standard RFS$_2$ in $\calO_4$,
$SR_2=(\bolda_1,\,\bolda_2; \, \zeta_2,\,\varphi_2)$:
\begin{eqnarray}
&& \bolda_1 \equiv s_1 s_2 ^* + s_3 s_4^*,           \label{uhf-rfs2-1}\\[-1pt]
&& \bolda_2 \equiv s_1 s_3 ^* - s_2 s_4^*,           \label{uhf-rfs2-2}\\[-1pt]
&& \zeta_2(X) \equiv s_1 X s_1^* - s_2 X s_2^* - s_3 X s_3^* + s_4 X s_4^*, 
                                                     \label{uhf-rfs2-3}\\[-3pt]
&& \varphi_2(X) \equiv \rho_4(X) 
               \equiv \sum_{i=1}^4 s_i X s_i^*,     \label{uhf-rfs2-4}\\[-15pt]
&&\nonumber
\end{eqnarray}
where $\rho_4$ is the canonical endomorphism of $\calO_4$.
The embedding $\varPhi_{SR_2}$ of CAR into $\calO_4$ associated with 
$SR_2$ is given by
\begin{eqnarray}
\noalign{\vskip-2pt}
&& \varPhi_{SR_2}(a_{2(n-1)+i}) = \zeta_2^{n-1}(\bolda_i), \quad 
     i=1,\,2,\, n=1,\,2,\,\ldots.                    \label{uhf-rfs2-5}
\end{eqnarray}
\end{list}
\vfill\eject\noindent
In the same way, the standard RFS$_p$ with a generic $p$ is explicitly
constructed.\cite{AK1}
\par
As for the standard RFS$_1$ in $\calO_2$, using mathematical induction, 
it is straightforward to see that $\varPhi_{SR_1}(\mbox{CAR})$ is identical 
with $\calO_2^{U(1)}$, which is defined by a linear space generated by 
monomials of the form 
$s_{i_1\cdots i_k;\,j_k\cdots j_1}$, $k=1,\,2,\,\ldots$.
Here, $\calO_2^{U(1)}$ is nothing but the $U(1)$-invariant subalgebra 
of $\calO_2$ with the $U(1)$ action been defined by $s_i \mapsto z s_i$,
$z\in\boldC$, $|z|=1$. From the one-to-one correspondence of 
$s_{i_1\cdots i_k;\, j_k \cdots j_1}$ with the matrix element 
$e_{i_1 j_1}\otimes\cdots\otimes e_{i_k j_k}\otimes I\otimes I\otimes\cdots$, 
$\calO_2^{U(1)}$ is isomorphic with 
$\bigotimes\limits^\infty M_2\cong\,$UHF$_2$, where $M_2$ denotes the algebra 
of all $2 \times 2$ complex matrices. In this correspondence, the embedding 
associated with $SR_1$ is transcribed into the form of infinite tensor 
products of matrices as follows:
\begin{alignat}{2}
\bolda &\sim  A\otimes I \otimes I \otimes \cdots, \quad & 
   A &\equiv \begin{pmatrix}0&1\\ 0&0\end{pmatrix}, \\[5pt]
\varPhi_{SR_1}(a_n) &\sim
 \underbrace{\sigma_3\otimes\cdots\otimes\sigma_3}_{n-1}\otimes A \otimes I
 \otimes I \otimes \cdots, \quad &
\sigma_3 &= \begin{pmatrix}1&0\\ 0&-1\end{pmatrix}.
\end{alignat}
Likewise, $\varPhi_{SR_p}(\mbox{CAR})$ is identical with $\calO_p^{U(1)}$.  
\par
Substituting the homogeneous embedding $\varPsi$ of $\calO_4$ into 
$\calO_2$ defined by\cite{AK2}
\begin{equation}
\begin{array}{c}
\varPsi : \calO_4 \hookrightarrow \calO_2, \\[5pt]
\varPsi(s_1)\equiv t_{11}, \quad
\varPsi(s_2)\equiv t_{21}, \quad
\varPsi(s_3)\equiv t_{12}, \quad
\varPsi(s_4)\equiv t_{22}, 
\end{array}                                          \label{O4-embeddedinO2}
\end{equation}
where $t_{ij}\equiv t_i\,t_j$ with $t_1$ and $t_2$ denoting the generator 
of $\calO_2$, into \eqref{uhf-rfs2-1}--\eqref{uhf-rfs2-3}, we obtain
\begin{eqnarray}
\varPsi(\bolda_1) 
&=& t_{11;\,12} + t_{12;\,22} = t_{1;\,2},           \label{uhf-rfs2-rfs1-1}\\
\varPsi(\bolda_2) 
&=& t_{11;\,21} - t_{21;\,22} = \zeta(\varPsi(\bolda_1)), 
                                                     \label{uhf-rfs2-rfs1-2}\\
(\varPsi\circ\zeta_2)(X) 
&=&  t_{11}\,\varPsi(X)\,t_{11}^* - t_{21}\,\varPsi(X)\,t_{21}^* 
   - t_{12}\,\varPsi(X)\,t_{12}^* + t_{22}\,\varPsi(X)\,t_{22}^* \nonumber\\
&=& \zeta^2(\varPsi(X)), \quad X \in \calO_4,        \label{uhf-rfs2-rfs1-3}\\
    \zeta(Y) &\equiv& t_1 Y t^*_1 - t_2Yt^*_2, \quad Y \in \calO_2,
                                                     \label{uhf-rfs2-rfs1-4}
\end{eqnarray}
where use has been made of \eqref{CR1} and \eqref{CR2} for $t_1$ and $t_2$.
Therefore, \eqref{uhf-rfs2-5} is rewritten as
\begin{eqnarray}
&&(\varPsi\circ\varPhi_{SR_2})(a_{2(n-1)+i}) 
 = \zeta^{2(n-1)+i-1}(\varPsi(\bolda_1)), \quad 
   i=1,\,2,\ \ n=1,\,2,\,\ldots, \\
\noalign{\noindent hence,}
&&(\varPsi\circ\varPhi_{SR_2})(a_n) 
 = \zeta^{n-1}(\varPsi(\bolda_1)), \quad n=1,\,2,\,\ldots,
\end{eqnarray}
which is nothing but the embedding $\varPhi_{SR}(a_n)$ associated with 
the standard RFS$_1$ in $\calO_2$ defined by \eqref{uhf-rfs-4}.
Likewise, the standard RFS$_p$ is reduced to the standard RFS$_1$ by the 
homogeneous embedding of $\calO_{2^p}$ into $\calO_2$.\cite{AK2}
\vskip10pt
\subsection{Representation}
As the $\ast$-representation of $\calO_d$, we consider the permutation 
representation.\cite{BJ} 
Let $\{e_n\, | \, n\in\boldN \}$ be an orthonormal basis of an 
infinite-dimensional Hilbert space $\calH$. 
Let $\mu_i: \boldN\to\boldN \ (i=1,\,\ldots,\,d)$ be a branching function
system defined by the following conditions: 
(i) 1-to-1, (ii) $\mu_i(\boldN)\cap\mu_j(\boldN)=\emptyset \ (i\not=j)$,
(iii) $\bigcup\limits_{i=1}^d \mu_i(\boldN)=\boldN$.
For a given branching function system $\{\mu_i\}$, the permutation 
representation of $\calO_d$ on $\calH$ is defined by
\begin{eqnarray}
s_i e_n = e_{\mu_i(n)}, \quad 
     i=1,\,2,\,\ldots,\,d, \ n=1,\,2,\,\ldots.       \label{PR1}
\end{eqnarray}
Here, we identify $s_i$ and its representation on $\calH$.
As for the action of $s_i^*$ on $\calH$, it is derived from the definition 
of the adjoint conjugation. Using the fact that any $n\in\boldN$ is uniquely 
expressed as $n=\mu_j(m)$ with appropriate $j$ and $m$, the result is 
written as 
\begin{eqnarray}
s_i^* e_n & = & s_i^* e_{\mu_j(m)} \nonumber\\
          & = & \delta_{i,j}e_m.                     \label{PR2}
\end{eqnarray}
An irreducible permutation representation is uniquely characterized by
a label $(i_1,i_2,\ldots,i_k)$ $(i_1,i_2,\ldots,i_k=1,2,\ldots,d)$ 
which has no periodicity less than $k$.  
Here, the label $(i_1,i_2,\ldots,i_k)$ is called to have periodicity 
$\ell(<k)$, if $i_1=i_{1+\ell}$, $i_2=i_{2+\ell},\,\ldots\,$, 
$i_{k-\ell}=i_k,\,i_{k-\ell+1}=i_1,\,\ldots\,$, $i_k=i_{\ell}$. 
Then, the irreducible permutation representation Rep$(i_1,\,\ldots,\,i_k)$
is defined by the case that the product $s_{i_1}\cdots s_{i_k}$ (and its
cyclic permutation) has the eigenvalue 1.  We set the corresponding
eigenvector of $s_{i_1}\cdots s_{i_k}$ on $e_1$.
Especially, for Rep$(1)$ or the standard representation of $\calO_d$, 
we have
\begin{alignat}{2} 
&s_i\, e_n = e_{d(n-1)+i}, 
 \quad && i=1,\,2,\, \ldots,\,d, \  n=1,\,2,\,\ldots, \\
&s_i^*\, e_{d(n-1)+j} = \delta_{i,j}e_n, 
 \quad && i,\,j=1,\,2,\,\ldots,\,d, \  n=1,\,2,\,\ldots. 
\end{alignat}
\par
Restricting Rep(1) of $\calO_2$ to the embedded image 
$\varPhi_{SR_1}(\mbox{CAR})$ associated with the standard RFS $SR_1$ 
in $\calO_2$, we obtain
\begin{eqnarray}
&&\varPhi_{SR_1}(a_n) \, e_1 
   = s_1^{n}s_2^*\,e_1=0, \qquad n=1,\,2,\,\ldots,   \label{rfs-rep1-1}\\
&&\varPhi_{SR_1}(a_{n_1})^* \varPhi_{SR_1}(a_{n_2})^* 
   \cdots \varPhi_{SR_1}(a_{n_k})^*\, e_1
  = s_1^{n_1-1}s_2s_1^{n_2-n_1-1}s_2\cdots s_1^{n_k-n_{k-1}-1}s_2\,e_1
       \qquad\nonumber\\
&&\phantom{\varPhi_{SR_1}(a_{n_1})^* \varPhi_{SR_1}(a_{n_2})^* 
   \cdots \varPhi_{SR_1}(a_{n_k})^*\, e_1}
  = e_{N(n_1,\ldots,n_k)}, \quad 1 \leqq n_1 <  \cdots  < n_k, \quad    
                       \label{rfs-rep1-2}\\
&&N(n_1,\ldots,n_k) \equiv 2^{n_1-1} + \cdots + 2^{n_k-1} + 1.
                                                     \label{rfs-rep1-3}
\end{eqnarray}
From \eqref{rfs-rep1-1}, $e_1$ of Rep(1) of $\calO_2$ is the vacuum for 
the annihilation operators $a_n$ $(n=1,\,2,\,\ldots)$. 
Since any $n\in\boldN$ is uniquely expressed as $N(n_1,\ldots,n_k)$ 
with $1\leqq n_1<\cdots<n_k$ in \eqref{rfs-rep1-3}, whole $\calH$ is 
interpreted as the Fock space with the unique vacuum $e_1$ for 
$\varPhi_{SR_1}(\mbox{CAR})$. 
\par
In the case of the standard RFS$_2$ in $\calO_4$, restricting Rep(1) of 
$\calO_4$, we obtain 
\begin{equation}
\varPhi_{SR_2}(a_{2(n-1)+i})\, e_1 = s_1^{n-1}\bolda_i\,e_1 = 0, 
    \quad i=1,\,2,\ n=1,\,2,\,\ldots.
\end{equation}
Thus, $e_1$ of Rep(1) of $\calO_4$ is a vacuum for the annihilation 
operators $\varPhi_{SR_2}(a_n)\ (n=1,\,2\,\ldots)$, and the corresponding 
Fock space is generated by 
$\varPhi_{SR_2}(a_{n_1})^*\varPhi_{SR_2}(a_{n_2})^*\cdots$ 
$\varPhi_{SR_2}(a_{n_k})^*\,e_1$ with $1\leqq n_1<n_2<\cdots<n_k$. 
As for 
$\varPhi_{SR_2}(a_{2(n_1-1)+i_1})^*\varPhi_{SR_2}(a_{2(n_2-1)+i_2})^*\cdots$
$\varPhi_{SR_2}(a_{2(n_k-1)+i_k})^* \, e_1$ with 
$1\leqq n_1<n_2<\cdots<n_k$ and $i_1,\,\ldots,\,i_k=1,\,2$, 
it is expressed in terms of a monomial consisting only of $s_1, \, s_2$ 
(for $i_j=1$) and $s_3$ (for $i_j=2$) acting on $e_1$.
On the other hand, in the case a product 
$\varPhi_{SR_2}(a_{2(n_j-1)+i_j})^*\varPhi_{SR_2}(a_{2(n_{j+1}-1)+i_{j+1}})^*$
with $n_j=n_{j+1}$ and $i_j=1$, $i_{j+1}=2$ is involved, it is expressed using
a monomial involving $s_4$.  
In this way, it is shown that
$\varPhi_{SR_2}(a_{2(n_1-1)+i_1})^*\cdots
\varPhi_{SR_2}(a_{2(n_k-1)+i_k})^*\,e_1$ 
with $1\leqq2(n_1-1)+i_1<\cdots<2(n_k-1)+i_k$ is expressed in the form of
\begin{eqnarray}
&&s_1^{n'_1-1}s_{i'_1} s_1^{n'_2-n'_1-1}s_{i'_2} \cdots 
     s_1^{n'_\ell-n'_{\ell-1}-1}s_{i'_{\ell}}e_1
   = e_{N(n'_1,\,i'_1;\,\ldots;\,n'_\ell,\,i'_\ell)},   
     \quad i'_1,\,\ldots,\,i'_\ell=2,\,3,\,4, \qquad\quad 
                                                     \label{rfs2-rep1-3}\\
&&N(n'_1,\,i'_1;\,\ldots;\,n'_\ell,\,i'_\ell)\equiv 
     \sum_{j=1}^\ell (i'_j-1) 4^{n'_j-1} +1,         \label{rfs2-rep1-4}
\end{eqnarray}
where $n_1= n'_1<\cdots<n'_\ell$, and $k-\ell$ is equal to the number of 
pairs of $\varPhi_{SR_2}(a_{2m-1})^*\varPhi_{SR_2}(a_{2m})$ in 
$\varPhi_{SR_2}(a_{2(n_1-1)+i_1})^*\cdots \varPhi_{SR_2}(a_{2(n_k-1)+i_k})^*$.
Since any $e_n\in\calH$ is uniquely expressed in the form of 
\eqref{rfs2-rep1-3}, whole $\calH$ is now interpreted as the Fock space 
with the unique vacuum $e_1$ for $\varPhi_{SR_2}(\mbox{CAR})$.
One should note that it is possible to rewrite the expression for 
$\varPhi_{SR_2}(a_{n_1})^*\varPhi_{SR_2}(a_{n_2})^*\cdots$ 
$\varPhi_{SR_2}(a_{n_k})^*\,e_1$ with $1\leqq n_1<n_2<\cdots<n_k$
into the same form as \eqref{rfs-rep1-2} with \eqref{rfs-rep1-3}.\cite{AK1}
Therefore, as a representation of CAR, the restriction of Rep(1) of $\calO_4$ 
to $\varPhi_{SR_2}(\mbox{CAR})$ is exactly the same as that of Rep(1) of 
$\calO_2$ to $\varPhi_{SR_1}(\mbox{CAR})$.
\vskip25pt
%
%
%
\section{Pseudo Cuntz Algebra}
We generalize the Cuntz Algebra $\calO_d$ defined on Hilbert spaces to the 
pseudo Cuntz algebra $\calO_{d,d'}$ on indefinite-metric vector 
spaces.\footnote{\samepage%
We only consider to generalize the $\ast$-algebraic structure of the Cuntz 
algebra, since it seems difficult to generalize the $C^*$-norm structure 
with mathematical rigorous.}
We consider a $\ast$-algebra generated by 
$s_1,\,\ldots,\,s_{d+d'}$ $(d+d'\geqq2)$ satisfying the following 
relations
\begin{eqnarray}
&& s_i^* s_j = \eta_{ij}I,                           \label{PCR1}\\
&& \sum_{i,j=1}^{d+d'}\eta^{ij}s_is_j^* = I,         \label{PCR2}
\end{eqnarray}
where $\eta_{ij}=\eta^{ij}=$diag$(\underbrace{+1,\,\cdots\,+1}_d,\,
\underbrace{-1,\,\cdots,\,-1}_{d'})$.  Then, $\calO_{d,0}$ is identical with
the dense subalgebra of $\calO_d$.
\par
For better understanding, let us introduce a vector space $\calV$ called 
the Krein space,\cite{Bognar} that is, a direct sum of two Hilbert spaces 
$\calV_\pm$, where $\calV_+$ has a positive definite inner product and 
$\calV_-$ has a negative definite one. 
We set the orthonormal basis $\{e_n\}\ (n=1,\,2,\,\ldots)$ of $\calV$ 
in such a way that $e_{2n-1}\in \calV_+$ and $e_{2n} \in \calV_-$.
Hence, we have
\begin{equation}
\langle\,e_m\,|\,e_n\,\rangle = (-1)^{m-1}\delta_{m,n}, \qquad
 m,\,n=1,\,2,\,\ldots,                               \label{indef-ip}
\end{equation}
where $\langle\, \cdot \, | \, \cdot \, \rangle$ denotes the inner product. 
We, next, define the operators $s_1$ and $s_2$ on $\calV$ by
\begin{eqnarray}
&&\begin{array}{lclclcl}
s_1 \, e_{2n-1} &=& e_{4n-3},  &\quad  &s_1\,e_{2n} &=& e_{4n},\\
s_2 \, e_{2n-1} &=& e_{4n-2},  &\quad  &s_2\,e_{2n} &=& e_{4n-1}, 
\end{array} 
\qquad n=1,\,2,\,\ldots,                             \label{pc-rep1}\\
&& \quad
s_1\ : \ \calV_\pm \to \calV_\pm, \qquad\qquad 
s_2\ : \ \calV_\pm \to \calV_\mp.                    \label{pc-map}
\end{eqnarray}
From the definition of the adjoint conjugation, the operation of $s_i^*$ 
on $e_n$ is uniquely derived from \eqref{indef-ip} and \eqref{pc-rep1} 
as follows:
\begin{equation}
\begin{array}{lclclcl}
s_1^* \, e_{4n-3} &=& e_{2n-1}, &\quad &s_1^* \, e_{4n-2} &=& 0, \\[5pt]
s_1^* \, e_{4n-1} &=& 0,        &\quad &s_1^* \, e_{4n} &=& e_{2n}, \\[5pt]
s_2^* \, e_{4n-3} &=& 0,        &\quad &s_2^* \, e_{4n-2} &=& -e_{2n-1},\\[5pt]
s_2^* \, e_{4n-1} &=& -e_{2n},  &\quad &s_1^* \, e_{4n} &=& 0, 
\end{array}
\qquad n=1,\,2,\,\ldots.                             \label{pc-rep1-star}
\end{equation}
Then, it is easy to see that $s_i$ and $s_j^*$ satisfy \eqref{PCR1} 
and \eqref{PCR2} with $d=d'=1$. 
The above $\ast$-representation, which we call Rep(1), of $\calO_{1,\,1}$ 
corresponds to Rep(1) of $\calO_2$ because $s_1$ has an eigenvector 
$e_1$ with eigenvalue 1, there is no other monomial $s_{i_1\cdots i_k}$
having eigenvector, and it is irreducible. In contrast to the case of 
$\calO_2$, there exists no representation such as Rep(2), in which $s_2$ has 
an eigenvector, because of \eqref{pc-map}. 
In general, a permutation representation $(i_i,\,i_2,\,\ldots,\,i_k)$ 
of $\calO_{1,\,1}$ is allowed only when the number of index 2 in 
$\{i_1,\,\ldots,\,i_k\}$ is even, since only in that case it is possible 
that $s_{i_1\cdots i_k}$ has an eigenvector.
\par
It is straightforward to construct permutation representations of 
$\calO_{d,\,d'}$. Here, we give the results for a special case of $d=d'$. 
In this case, it is convenient to rearrange the generators of $\calO_{d,d}$ 
in such a way that $\eta_{ij}$ in \eqref{PCR1} and \eqref{PCR2} is given by 
$\eta_{ij}=(-1)^{i-1}\delta_{ij}$ $(i,\,j=1,\,\ldots,\,2d)$.
Then, Rep(1) of $\calO_{d,d}$ on the Krein space $\calV$ is given by
\begin{alignat}{3}
s_i\,e_{2n-1} &= e_{4d(n-1)+i}, &\quad
s_i\,e_{2n}   &= e_{4dn + 1 -i}, &\quad
              & i=1,\,\ldots,\,2d,\\
s_i^*\,e_{4d(n-1)+j} &= (-1)^{i-1}\delta_{i,j}\,e_{2n-1}, &\quad
s_i^*\,e_{4dn+1-j}   &= (-1)^{i-1}\delta_{i,j}\,e_{2n},    &\quad
                     & i,\,j=1,\,\ldots,\,2d,\\\
s_{2j-1}&\,: \ \calV_{\pm} \to \calV_{\pm}, &\quad
  s_{2j}&\,: \ \calV_{\pm} \to \calV_{\mp}, &\quad
        & j=1,\,\ldots,\,d.
\end{alignat}
\vskip30pt
%
%
%
\section{Recursive FP Ghost System in String Theory}
We denote the $\ast$-algebra generated by the FP (anti)ghosts in string 
theory by $\calFP$.
The generators of $\calFP$ are FP ghost $c_n$ and FP antighost 
$\barc_n$ $(n=0,\,1,\,2,\,\ldots)$ with $c_0{}^*=c_0$ and 
$\barc_0{}^*=\barc_0$.
They satisfy the following anticommutation relations
\begin{eqnarray}
&&\{ c_0, \,\barc_0 \} = -I,                         \label{fp0} \\ 
&&\{ c_m,\,\barc_n^* \} = -\delta_{m,n}\,I, \qquad m,\,n=1,\,2,\,\ldots,
                                                     \label{fp1}\\
&&\{ c_m,\,c_n \}=\{ c_m,\,c_n^* \}
 =\{ \barc_m,\,\barc_n \}=\{ \barc_m,\,\barc_n^* \}=0,
  \quad m,\,n=0,\,1,\,\ldots,                        \label{fp2} \\ 
&&\{ c_0,\,\barc_n \} = \{ c_m,\,\barc_0 \}
 =\{ c_m,\,\barc_n \}=0, \qquad m,\,n=1,\,2,\,\ldots.\label{fp3} 
\end{eqnarray} 
The purpose of this section is to give the recursive construction for 
embedding of $\calFP$ into $\calO_{2,\,2}$.
\par
First, we introduce ICAR defined by the fermion algebra with indefinite 
signature, in which the generators $a_n$ $(n=1,\,2,\,\ldots)$ satisfy
\begin{equation}
\{ a_m, \, a_n \} = 0, \quad \{ a_m, \, a_n^* \} = (-1)^m \delta_{m,n}I,
\quad m,\,n=1,\,2,\,\ldots.                          \label{indefinite-car}
\end{equation}
We construct an embedding of ICAR into $\calO_{2,2}$ with 
$\eta_{ij}=(-1)^{i-1}\delta_{ij}$ by generalizing RFS$_2$ in $\calO_4$.
Let $\bolda_1,\,\bolda_2\in\calO_{2,2}$, 
$\zeta_{1+1}: \calO_{2,2}\to\calO_{2,2}$ be a linear mapping, 
and $\varphi_{1+1}$ a unital endomorphism of $\calO_{2,2}$, respectively.
A tetrad $R_{1+1}=(\bolda_1,\,\bolda_2;\,\zeta,\,\varphi)$ is called the 
{\it recursive fermion system of $(1,1)$-type\/} (RFS$_{1+1}$) in 
$\calO_{2,2}$, if it satisfies\footnote{In $\calO_{2,2}$, there exists 
also the anticommutation relation algebra with negative-definite 
signature as will be shown in the last section.}
\begin{alignat}{3}
&\mbox{$\phantom{\mbox{ii}}$i) seed condition:}&\ \ &
 \{\bolda_i, \, \bolda_j\} \!=\! 0, \quad
 \{\bolda_i, \, \bolda_j^*\}\!=\!(-1)^i\delta_{i,j}I, \quad i,\,j=1,\,2, 
                                                     \label{rfs1,1-1}\\
&\mbox{$\phantom{\mbox{i}}$ii) recursive condition:}&\ \ &
\{\bolda_i,\,\zeta_{1+1}(X)\}\!=\! 0, \  
\zeta_{1+1}(X)^*\!=\!\zeta_{1+1}(X^*), \ \ 
  X \in \calO_{2,2},           \label{rfs1,1-2}\\
&\mbox{iii) normalization condition:}&\ \ &
\zeta_{1+1}(X)\zeta_{1+1}(Y)\!=\!\varphi_{1+1}(XY), \quad 
  X,\,Y \in \calO_{2,2}.       \label{rfs1,1-3}
\end{alignat}
Then, the embedding $\varPhi_{R_{1+1}}$ of ICAR into $\calO_{2,2}$ 
associated with $R_{1+1}$ is defined by
\begin{equation}
\begin{array}{c}
\varPhi_{R_{1+1}} : \mbox{ICAR} \hookrightarrow \calO_{2,2}, \\[5pt]
\varPhi_{R_{1+1}}(a_{2(n-1)+i} )\equiv \zeta_{1+1}^{n-1}(\bolda_i), \quad 
              i=1,\,2,\ \ n=1,\,2,\,\ldots.
\end{array}                                          \label{rfs1,1-4}
\end{equation}
It is straightforward to reconfirm that \eqref{rfs1,1-4} satisfy the 
anticommutation relation of \eqref{indefinite-car}.
The simplest example of RFS$_{1+1}$ in $\calO_{2,2}$ is given by 
the {\it standard\/} RFS$_{1+1}$,  $SR_{1+1}=(\bolda_1, \, 
\bolda_2;\,\zeta_{1+1}, \, \varphi_{1+1})$, which is defined by
\begin{eqnarray}
&&\bolda_1 \equiv s_1 s^*_2 + s_3 s^*_4, \quad 
  \bolda_2 \equiv s_1 s^*_3 + s_2 s^*_4, 
                                                     \label{uhf-rfs1,1-1}\\
&&\zeta_{1+1}(X) \equiv s_1 X s_1^* + s_2 X s_2^* - s_3 X s_3^* - s_4 X s_4^*, 
                                                   \label{uhf-rfs1,1-2}\\[-3pt]
&&\varphi_{1+1}(X) \equiv \rho_{2,2}(X) 
  \equiv \sum_{i=1}^4 (-1)^{i-1}s_i X s_i^*,      \label{uhf-rfs1,1-3}\\[-18pt]
&&\nonumber
\end{eqnarray}
where $\rho_{2,2}$ should be called the canonical endomorphism of 
$\calO_{2,2}$.
Like the standard RFS$_2$ in $\calO_4$, $\varPhi_{SR_{1+1}}(\mbox{ICAR})$
is identical with the $\calO_{2,2}^{U(1)} \subset \calO_{2,2}$, which is 
a linear space spanned by monomials of the form 
$s_{i_1\cdots i_k;\,j_k\cdots j_1}$ $(k=1,\,2,\,\ldots)$.
\vskip5pt
\subsection{RFPS1}
As noted in Sec.$\,$1, the subalgebra generated by the positive-mode operators 
of FP (anti)ghost $c_n,\,\barc_n$ $(n=1,\,2,\,\ldots)$ is isomorphic with 
ICAR, while the subalgebra generated by the zero-mode operators $c_0$ and 
$\barc_0$ is isomorphic with a $(1+1)$-dimensional Clifford algebra.
The generators of the $(1+1)$-dimensional Clifford algebra $b_1$ and $b_2$ 
are written in terms of those in the $2$-dimensional ICAR as follows
\begin{eqnarray}
\noalign{\vskip-2pt}
&&b_i = b_i^* = a_i + a_i^*, \quad i=1,\,2,          \label{clifford-1}\\[-2pt]
&&\{ b_i, \, b_j \} =2\eta_{i,j}I,  
  \quad \eta_{i,j}=\mbox{diag}(-1,\, +1),            \label{clifford-2}
\end{eqnarray}
where $a_1$ and $a_2$ satisfy \eqref{indefinite-car}.
Therefore, we have a natural correspondence between generators of 
$\calFP$ and those of ICAR as follows:
\begin{eqnarray}
\noalign{\vskip-2pt}
&&c_0 = \frac{1}{2}(b_1+b_2)
      =\frac{1}{2}(a_1 + a_2 + a_1^* + a_2^*),       \label{fp-icar1}\\
&&\barc_0 = \frac{1}{2}(b_1-b_2)
          = \frac{1}{2}(a_1 - a_2 + a_1^* - a_2^*),  \label{fp-icar2}\\
&&c_n = \frac{1}{\sqrt2}(a_{2n+1}+a_{2n+2}), \quad n=1,\,2,\,\ldots,
                                                     \label{fp-icar3}\\[-5pt]
\noalign{\eject}
&&\barc_n = \frac{1}{\sqrt2}(a_{2n+1}-a_{2n+2}), \quad n=1,\,2,\,\ldots, 
                                                     \label{fp-icar4}
\end{eqnarray}
where $a_n$ $(n=1,\,2,\,\ldots)$ satisfy \eqref{indefinite-car}.
From \eqref{rfs1,1-4} and \eqref{fp-icar1}--\eqref{fp-icar4}, it is 
straightforward to obtain the embedding $\varPhi_{RFP1}$ of $\calFP$ 
into $\calO_{2,2}$ defined by
\begin{alignat}{3}
\varPhi_{RFP_1} &: \calFP \hookrightarrow \calO_{2,2},  && \nonumber\\
\varPhi_{RFP1}(c_0) &\equiv \frac{1}{\sqrt2}(\boldc + \boldc^*), &\quad 
\varPhi_{RFP1}(\barc_0) &\equiv \frac{1}{\sqrt2}(\boldbarc + \boldbarc^*), &&
                                                     \label{rfps-1}\\[2pt]
\varPhi_{RFP1}(c_n) &\equiv \zeta_{1+1}^n(\boldc), &\quad
\varPhi_{RFP1}(\bar c_n) &\equiv \zeta_{1+1}^n(\boldbarc), &
\quad& n=1,\,2,\,\ldots,                           \label{rfps-2}\\[2pt]
\boldc &\equiv \frac{1}{\sqrt2}(\bolda_1 + \bolda_2), &\quad
\boldbarc &\equiv \frac{1}{\sqrt2}(\bolda_1-\bolda_2),&&\label{rfps-3}
\end{alignat}
where $\boldc$ and $\boldbarc$ satisfy
\begin{eqnarray}
\noalign{\vskip-5pt}
&& \boldc^2 = \boldbarc^2 = \{ \boldc, \, \boldbarc \}= 0, \\
&& \{ \boldc, \, \boldbarc^* \} = - I.
\end{eqnarray} 
Substituting $\bolda_1$ and $\bolda_2$ of the standard RFS$_{1+1}$ 
defined by \eqref{uhf-rfs1,1-1} into \eqref{rfps-3} and \eqref{rfps-1}, 
we obtain
\begin{eqnarray}
\noalign{\vskip-5pt}
&& \boldc = \frac{1}{\sqrt2}[ s_1(s_2^* + s_3^*) + (s_2 + s_3)s_4^* ], 
                                                     \label{c}\\ [2pt]
&& \boldbarc = \frac{1}{\sqrt2}[ s_1(s_2^* - s_3^*) - (s_2 - s_3)s_4^* ], 
                                                     \label{barc}\\ [2pt]
&& \varPhi_{RFP1}(c_0) = \frac{1}{2}[ (s_1 + s_4)(s_2^* + s_3^*) 
                      + (s_2 + s_3)(s_1^* + s_4^*) ],\label{c0-1} \\ [2pt]
&& \varPhi_{RFP1}(\barc_0) = \frac{1}{2}[ (s_1 - s_4)(s_2^* - s_3^*) 
                      + (s_2 - s_3)(s_1^* - s_4^*) ].\label{barc0-1}
\end{eqnarray}
We call the tetrad $RFP1=(\boldc,\,\boldbarc;\,\zeta_{1+1},\,\varphi_{1+1})$ 
the {\it recursive FP ghost system of the first type (RFPS1).}
\vspace{10pt}
\subsection{RFPS2}
The embedding of $\calFP$ into $\calO_{2,2}$ is not uniquely given 
by RFPS1. 
To construct another one, let us note the existence of an embedding of 
the $(1+1)$-dimensional Clifford algebra generated by $b'_1$ and $b'_2$ 
into the $1$-dimensional CAR with negative-definite signature, which is 
given by
\begin{eqnarray}
b'_1 &\equiv& a_1 + a_1^*,                  \label{anotherclifford-1} \\
b'_2 &\equiv& \exp(i\pi a_1^* a_1) 
  = I + 2a_1^* a_1= a_1^* a_1 - a_1 a_1^*,  \label{anotherclifford-2}
\end{eqnarray}
where $a_1$ constitutes a 1-dimensional $\ast$-subalgebra of ICAR 
\eqref{indefinite-car}.
Here, $\exp(i\pi a_1^* a_1)$ is the Klein operator anticommuting with 
$a_1$. One should note that an identity $\exp(i2\pi a_1^* a_1)=I$ holds. 
Substituting the expressions for $\bolda_1$ of the standard RFS$_{1,1}$ 
defined by \eqref{uhf-rfs1,1-1} into $a_1$ in \eqref{anotherclifford-1} 
and \eqref{anotherclifford-2}, we obtain another embedding $\varPhi_{RFP2}$ 
of  the zero-mode operators $c_0$ and $\barc_0$ into $\calO_{2,2}$ 
as follows:
\vfill\eject
\begin{eqnarray}
\varPhi_{RFP2} &: &  \calFP \hookrightarrow \calO_{2,2}, \nonumber\\[5pt]
\varPhi_{RFP2}(c_0) 
  &\equiv& \frac{1}{2}(\bolda_1 + \bolda^*_1 
           + \bolda_1^* \bolda_1 - \bolda_1 \bolda^*_1) \nonumber\\[2pt]
  &=& \frac{1}{2}[ (s_1 + s_2)(s_1^* + s_2^*) + (s_3 + s_4)(s_3^* + s_4^*) ], 
                                                     \label{c0-2}\\[2pt]
\varPhi_{RFS2}(\barc_0) 
  &\equiv& \frac{1}{2}(\bolda_1 + \bolda^*_1 
           - \bolda_1^* \bolda_1 + \bolda_1 \bolda^*_1) \nonumber\\[2pt]
  &=& -\frac{1}{2}[ (s_1 - s_2)(s_1^* - s_2^*) + (s_3 - s_4)(s_3^* - s_4^*) ].
                                                     \label{barc0-2}
\end{eqnarray}
In contrast with $b_2$ defined by \eqref{clifford-1} with $i=2$, $b'_2$ 
no longer anticommutes with $a_n$ $(n=3,\,4,\,\ldots)$, but commutes 
with them. In other words, since \eqref{anotherclifford-2} is nonlinear 
in $a_1$, $\zeta_{1+1}(X)$ no longer anticommutes with it, hence the 
previous embedding $\varPhi_{RFS1}(c_n)$ and $\varPhi_{RFS1}(\barc_n)$ 
$(n=1,\,2,\,\ldots)$ do not anticommutes with $\varPhi_{RFS2}(c_0)$ 
and $\varPhi_{RFS2}(\barc_0)$.
In order to recover the anticommutativity, we introduce a new mapping 
$\zeta_0 : \calO_{2,2} \to \calO_{2,2}$ defined by
\begin{equation}
\zeta_0(X) \equiv s_2 X s_1^* - s_1 X s_2^* + s_4 X s_3^* - s_3 X s_4^*,
 \quad X \in \calO_{2,2}, 
                                                     \label{zeta0}
\end{equation}
which satisfies
\begin{eqnarray}
&& \{ \varPhi_{RFP2}(c_0), \, \zeta_0(X)\} = 
   \{ \varPhi_{RFP2}(\barc_0),\,\zeta_0(X)\} = 0, 
   \quad \zeta_0(X)^* =  -\zeta_0(X^*),              \label{c0-zeta0}\\
&& \zeta_0(X)\zeta_0(Y) = \varphi_{1+1}(XY),
\end{eqnarray}
where $\varphi_{1+1}$ is defined by \eqref{uhf-rfs1,1-3}. 
Since $\zeta_0(I)=s_{2;\,1}-s_{1;\,2}+s_{4;\,3}-s_{3;\,4}
=\bolda^*_1-\bolda_1$, the anticommutativity in \eqref{c0-zeta0} 
is owing to $\{a_1 + a_1^*, \, a_1^* - a_1\}$
$=\{\exp(i\pi a_1^* a_1), \, a_1^* - a_1 \}=0$.
Then, we define another embedding $\varPhi_{RFP2}$ of the positive-mode 
operators $c_n$ and $\barc_n$ as follows
\begin{equation}
\varPhi_{RFP2}(c_n) \equiv \zeta_0(\zeta_{1+1}^{n-1}(\boldc)), \quad 
\varPhi_{RFP2}(\barc_n) \equiv -\zeta_0(\zeta_{1+1}^{n-1}(\boldbarc)),
 \qquad n=1,\,2,\,\ldots,                            \label{another-cn-cbarn}
\end{equation}
where $\boldc$, $\boldbarc$ and $\zeta_{1+1}$ are defined by \eqref{c}, 
\eqref{barc} and \eqref{uhf-rfs1,1-2}, respectively.
It is straightforward to show that the above $\varPhi_{RFP2}(c_n)$ and 
$\varPhi_{RFP2}(\barc_n)$ $(n\geqq0)$ indeed satisfy \eqref{fp0}--\eqref{fp3}. 
We call a set 
$RFP2=(c,\,\barc,\,c_0,\,\barc_0;\, \zeta_{1+1},\,\zeta_0,\,\varphi_{1+1})$ 
the {\it recursive FP ghost system of the second type (RFPS2).}
\par
The apparent difference between RFPS1 and RFPS2 is only that the degree 
of freedom corresponding to the generator $a_2$ of ICAR disappears in 
the latter.
As shown in the next section, the most significant difference of them 
is that they correspond to two unitarily inequivalent representations 
of $\calFP$.
\vskip30pt
%
%
%
\section{Representation of RFPS}
In this section, we consider restrictions of Rep(1) of $\calO_{2,2}$ 
to $\varPhi_{RFP1}(\calFP)$ and $\varPhi_{RFP2}(\calFP)$.
First, we recall the Rep(1) of $\calO_{2,2}$ on the Krein space $\calV$:
\vfill\eject
\begin{alignat}{2}
s_i\, e_{2n-1} &= e_{8(n-1)+i}, &\quad 
s_i\,e_{2n} &= e_{8n+1- i},\\
s_i^*\,e_{8(n-1)+j} &= (-1)^{i-1}\delta_{i,j}\,e_{2n-1}, &\quad
s_i^*\,e_{8n+1-j} &= (-1)^{i-1}\delta_{i,j}\,e_{2n},  
\end{alignat}
where $i,\,j=1,\,2,\,3,\,4; \ n=1,\,2,\,\ldots$.  
\vskip10pt
\subsection{Restriction of Rep(1) to RFPS1}
In this subsection, we denote $C_n\equiv\varPhi_{RFP1}(c_n)$, 
$\barC_n\equiv\varPhi_{RFP1}(\barc_n)$ $(n\geqq0)$ for simplicity of 
description.
\par
Since $\zeta_{1,1}$ defined by \eqref{uhf-rfs1,1-2} satisfies
\begin{equation}
\zeta_{1+1}(X)s_i= \epsilon_i s_i X, \quad 
\epsilon_i\equiv(-1)^{i-1+\left[\frac{i-1}{2}\right]},
\quad i=1,\,2,\,3,\,4,
\end{equation}
where $[x]$ denotes the largest integer not greater than $x$, we obtain
\begin{equation}
C_n\,s_i\,e_1=\epsilon_i s_i s_1^{n-1} \boldc\,e_1=0, \quad
\barC_n\,s_i\,e_1=\epsilon_i s_i s_1^{n-1} \boldbarc\,e_1=0, 
\quad n=1,\,2,\,\ldots.
\end{equation}
Here, use has been made of $s_1^*\,e_1=e_1$, $s_i^*\,e_1=0$ $(i=2,\,3,\,4)$ 
and
\begin{equation}
\boldc \, e_1 = \boldbarc \, e_1 =0.
\end{equation}
Hence, we have at least four vacuums $s_i\,e_1$ with respect to the 
positive-mode operators. 
In fact, there is no other vacuums annihilated by $C_n$ and 
$\barC_n$ $(n=1,\,2,\,\ldots)$ 
because of the cyclicity of the representation as shown later.
\par
As for the zero-mode operators defined by \eqref{c0-1} and \eqref{barc0-1},  
we have
\begin{eqnarray}
&& C_0 \, s_1 = \frac{1}{2}(s_2 + s_3), \\
&& \barC_0 \, s_1 = \frac{1}{2}(s_2 - s_3),\\
\noalign{\noindent hence,}
&& s_2 = ( C_0 + \barC_0 )\,s_1,                     \label{s_2-C-barC}\\
&& s_3 = ( C_0 - \barC_0 )\,s_1.                     \label{s_3-C-barC}
\end{eqnarray}
Furthermore, since $C_0$ and $\barC_0$ satisfy
\begin{equation}
[ \barC_0,\, C_0] = s_{4,1}-s_{1,4}+s_{2,3}-s_{3,2},
\end{equation}
we have
\begin{equation}
s_4 = [ \barC_0,\, C_0 ]\,s_1.                       \label{s_4-C-barC}
\end{equation}
Thus, $\{s_1,\,s_2,\,s_3,\,s_4\}$ can be interpreted as a four-dimensional 
representation space of the zero-mode operators.
Therefore, the four vacuums $s_i e_1$ with respect to the positive-mode 
operators are expressed by the zero-mode operators and $e_1$.
From the anticommutativity between the positive-mode operators and 
the zero-mode operators, in order to construct Fock space based on 
the above four vacuums, it is sufficient to consider the action of 
positive-mode creation operators on $e_1$ only.
\par
Because of the anticommutativity between two creation operators of 
different modes, we only have to consider 
$\varphi_{n_1}^*\cdots\varphi_{n_k}^*\,e_1$ with $\varphi_{2m-1}
\equiv\barC_m$, $\varphi_{2m}\equiv C_m$, $1\leqq n_1<\cdots<n_k$. 
If $\varphi_{n_1}^*\cdots\varphi_{n_k}^*$ does not involve the product 
$\barC_m^* C_m^*$ (i.e., $n_{i+1}>n_i+1$ for odd $n_i$), we obtain the 
following
\begin{eqnarray}
&&\hspace*{-20pt}
\varphi_{n_1}^* \varphi_{n_2}^* \cdots \varphi_{n_k}^*\,e_1 
= \frac{1}{\sqrt{2^k}}
   s_1^{m_1}(s_2+(-1)^{n_1}s_3)
   s_1^{m_2-m_1-1}(s_2+(-1)^{n_2}s_3) 
   \cdots  \nonumber\\[-5pt]
&& \hspace*{150pt}\cdots
   s_1^{m_k-m_{k-1}-1}(s_2+(-1)^{n_k}s_3)\,e_1,      \label{phi_k-1} 
\end{eqnarray}
where $m_i\equiv\left[\frac{n_i+1}{2}\right]$ with $[\,x\,]$ denoting 
the largest integer not greater than $x$.
When $\barC_{m_j}^* C_{m_j}^*$ is involved in 
$\varphi_{n_1}^*\cdots\varphi_{n_k}^*$ (i.e., $n_{i+1}=n_i+1$ for some odd 
$n_i$) in \eqref{phi_k-1},  the corresponding factors in rhs are replaced 
by $2s_1^{m_j-\cdots}s_4$. For example, we have
\begin{eqnarray}
&&\hspace*{-20pt} 
\barC_m^* C_m^*\, e_1 = s_1^m s_4\,e_1,\\
&&\hspace*{-20pt} 
C_{m_1}^* \cdots \barC_{m_j}^* C_{m_j}^* \cdots C_{m_k}^*\,e_1 
= \frac{1}{\sqrt{2^{k-1}}}s_1^{m_1}(s_2+s_3)
            s_1^{m_2-m_1-1}(s_2+s_3) \cdots\nonumber\\[-5pt]
&& \hspace*{150pt}   
   \cdots s_1^{m_j-m_{j-1}-1}s_4 \cdots 
          s_1^{m_k-m_{k-1}-1}(s_2+s_3)\,e_1.
\end{eqnarray}
Thus, the vector space $\calV^{(0)}$ generated by action of the 
positive-mode creation operators on $e_1$ is a linear space spanned 
by $s_1^{n_1}s_{i_1}s_1^{n_2-n_1-1}s_{i_2}\cdots 
s_1^{n_\ell-n_{\ell-1}-1}s_{i_\ell}\,e_1$ with
$\ell\in\boldN$, $1\leqq n_1<n_2<\cdots<n_\ell$ and 
$i_1,\,\ldots,\,i_\ell=2,\,3,\,4$.
Since any $e_n\in\calV$ is obtained by action of an appropriate 
monomial consisting of $s_i\ (i=1,\,2,\,3,\,4)$ on $e_1$, and such 
a monomial is uniquely expressed by 
$s_1^{n_1-1}s_{i_1}s_1^{n_2-n_1-1}s_{i_2}\cdots 
s_1^{n_\ell-n_{\ell-1}-1}s_{i_\ell}$ 
$(\ell\in\boldN$, $1\leqq n_1<n_2<\cdots<n_\ell$; 
$i_1,\,\ldots,\,i_\ell=2,\,3,\,4)$, 
the above results show that $\calV^{(0)} = s_1\calV$.
Therefore, taking the contribution from the zero-mode operators into account, 
we obtain
\begin{eqnarray}
\calV
&=&s_1\calV \oplus s_2\calV \oplus s_3\calV \oplus s_4 \calV \nonumber\\
&=&\calV^{(0)} \oplus C_0\calV^{(0)} 
               \oplus \barC_0 \calV^{(0)} 
               \oplus [ \barC_0,\, C_0 ] \calV^{(0)},\label{V}
\end{eqnarray}
where use has been made of \eqref{s_2-C-barC}, \eqref{s_3-C-barC} and 
\eqref{s_4-C-barC}.
Hence, the restriction of Rep(1) of $\calO_{2,2}$ to 
$\varPhi_{RFP1}(\calFP)$ is cyclic with the cyclic vector $e_1$.
This type of representation of the FP ghost algebra in string theory 
was found through the exact Wightman functions in the operator formalism 
by one of the present authors (M.A.) and Nakanishi.\cite{AN}
\par
From \eqref{V}, we can express the subspace $\calV^{(0)}$ irrelevant
to the zero-mode FP (anti)ghost operators as follows
\begin{eqnarray}
\calV^{(0)}
 &=& s_1 s_1^*\, \calV \nonumber\\
 &=&\{ v\in \calV\, | \, s_i^*v =0, \ i=2,\,3,\,4\},
\end{eqnarray}
where the first line shows that $s_1 s_1^*$ is the projection operator to 
$\calV^{(0)}$ and the second one denotes the subsidiary condition to 
select $\calV^{(0)}$. 
\vspace{10pt}
\subsection{Restriction of Rep(1) to RFPS2}
From \eqref{another-cn-cbarn}, it is straightforward to have
\begin{equation}
\varPhi_{RFP2}(c_n)s_i\,e_1 = \varPhi_{RFP2}(\barc_n)s_i\,e_1 =0, \quad
i=1,\,2,\,3,\,4,\ \ n=1,\,2,\,\ldots
\end{equation}
in the same way as in RFPS1.  Therefore, $s_i\,e_1$'s are vacuums for 
the positive-mode operators also in this case.
To specify the contribution from the zero-mode operators, we consider 
their action on $s_i$. From \eqref{c0-2} and \eqref{barc0-2}, it is easy 
to have
\begin{equation}
\varPhi_{RFP2}(c_0)\,(s_{2i-1}- s_{2i})= s_{2i-1} + s_{2i},        \ \ 
\varPhi_{RFP2}(\barc_0)\,(s_{2i-1}+s_{2i})= - (s_{2i-1} - s_{2i}), \ \ 
i=1,\,2.
\end{equation}
Thus, in contrast with RFPS1, $\{s_1,\,s_2,\,s_3,\,s_4\}$ is a direct sum 
of two two-dimensional representations of the zero-mode operators.
Based on this features of RFPS2 and the similar consideration on the 
positive-mode operators in RFPS1, it is straightforward to obtain that 
the total space $\calV$ is expressed as follows
\begin{alignat}{2}
&\calV = \calV_1 \oplus \calV_2, &\quad 
&\calV_1 \perp \calV_2,\\
&\calV_i \equiv \calV_i^{(+)}\oplus\calV_i^{(-)}, &\quad
&\calV_i^{(\pm)} \equiv \frac{s_{2i-1}\pm s_{2i}}{\sqrt2}\calV, 
 \quad i=1,\,2, \\
&\calV_i^{(+)} = \varPhi_{RFP2}(c_0) \calV_i^{(-)}, &\quad 
&\calV_i^{(-)} = - \varPhi_{RFP2}(\barc_0) \calV_i^{(+)}, 
\end{alignat}
where $\calV_i^{(\pm)}$ is a Fock space with the vacuum 
$e^{(\pm)}_i\equiv\frac{s_{2i-1}\pm s_{2i}}{\sqrt{2}}e_1$ 
for the positive-mode operators. Therefore, each $\calV_i$ is an 
invariant subspace for $\varPhi_{RFP2}(\calFP)$.
Here,  $e^{(\pm)}_i$ satisfy the following:
\begin{alignat}{3}
&e^{(+)}_i = \varPhi_{RFP2}(c_0) \, e^{(-)}_i, &\quad
&e^{(-)}_i = -\varPhi_{RFP2}(\barc_0) \, e^{(+)}_i, &\quad 
&i=1,\,2,\\
&\langle\, e^{(\pm)}_i\,|\,e^{(\mp)}_j\,\rangle = \delta_{i,j}, &\quad
&\langle\, e^{(\pm)}_i\,|\,e^{(\pm)}_j\,\rangle = 0, &\quad 
&i,\,j=1,\,2.                                        \label{KO-vac}
\end{alignat}
Therefore, $e^{(\pm)}_i$ for each $i$ correspond to the two-dimensional
vacuums $|\,\pm\,\rangle$ introduced by Kato and Ogawa.\cite{KO}
Since each $e^{(\pm)}_i$ is orthogonal with itself, the naive definition 
of vacuum expectation value is not appropriate in this representation of 
$\calFP$. In order to recover the vacuum expectation value in the 
conventional sense, one is apt to introduce a metric operator $\eta$ 
satisfying $\eta\,e^{(\pm)}_i=e^{(\mp)}_i$.
It should be noted, however, that the original definition of 
$\ast$-involution (or Hermitian conjugation) in the $\ast$-algebra 
is {\it not\/}, in general, respected in such an expectation value 
defined in terms of the metric operator.
In the physical point of view, there is no reason to adhere to this kind 
of representation of $\calFP$, in which there is {\it no\/} vacuum 
(or cyclic vector) having positive norm.
\vskip30pt
%
%
%
\section{Discussion}
In the present paper, we have introduced the pseudo Cuntz algebra and
constructed two recursive FP ghost systems, RFPS1 and RFPS2 in $\calO_{2,2}$, 
and their representations. As shown in the Appendix, there exists an embedding 
$\hat\varPsi$ of $\calO_{2,2}$ into $\calO_{1,1}$ such as
\begin{equation}
\begin{array}{c}
\hat\varPsi : \calO_{2,2} \hookrightarrow \calO_{1,1}, \\[5pt]
\hat\varPsi(s_1)\equiv t_{11}, \quad 
\hat\varPsi(s_2)\equiv t_{21}, \quad 
\hat\varPsi(s_3)\equiv t_{22}, \quad 
\hat\varPsi(s_4)\equiv t_{12}, 
\end{array}
                                                     \label{embed2,2-1,1}
\end{equation}
where $t_{ij}\equiv t_i\,t_j$ with $t_1$ and $t_2$ denoting generators 
of $\calO_{1,1}$, hence, it is, of course, possible to discuss on the 
recursive FP ghost system in $\calO_{1,1}$. 
However, we dare not to have done so because it would be rather 
complicated and not adequate to make clear-cut descriptions.
Indeed, substituting \eqref{embed2,2-1,1} into \eqref{uhf-rfs1,1-1} and
\eqref{uhf-rfs1,1-2}, the resultant expressions would not be so simple as
\eqref{uhf-rfs2-rfs1-1}--\eqref{uhf-rfs2-rfs1-4}.  
This is because the natural fermion subalgebra in $\calO_{1,1}$ is 
not ICAR but NCAR considered bellow. In contrast with $\calO_{1,1}$, we can 
treat embeddings of ICAR and NCAR in parallel in $\calO_{2,2}$.
Hence, it seems most transparent to discuss RFPS in $\calO_{2,2}$.  
\par
As for the recursive fermion systems in the pseudo Cuntz algebra, 
there is a very impressive phenomenon as follows.  
We consider a fermion algebra NCAR with negative-definite signature, in which 
generators $a'_n$ $(n=1,\,2,\,\ldots\,)$ satisfy the anticommutation 
relations as follows
\begin{equation}
\{a'_m, \, a'_n\}=0, \quad
\{a'_m, \, a^{\prime\,\ast}_n\} = - \delta_{m,n}I, \quad 
  m,\,n=1,\,2,\,\ldots.
\end{equation}
It is possible to embed NCAR into $\calO_{2,2}$ in the following way.
Let us define a tetrad $SR_{0+2}=(\bolda'_1, \, \bolda'_2;\,\zeta_{0+2}, 
\varphi_{0+2})$, which is called the {\it standard recursive fermion system
of $(0,2)$-type\/} (standard RFS$_{0+2}$) in $\calO_{2,2}$, by\footnote{%
Substituting \eqref{embed2,2-1,1} into 
\eqref{uhf-rfs0,2-1}--\eqref{uhf-rfs0,2-4}, we obtain a RFS in 
$\calO_{1,1}$ in the similar form of $SR_1$ in $\calO_2$, which should 
be called the standard RFS$_{0+1}$ since it gives an embedding of NCAR 
onto $\calO_{1,1}^{U(1)}$.}
\begin{eqnarray}
&& \bolda'_1 \equiv s_1 s^*_2 - s_4 s^*_3,           \label{uhf-rfs0,2-1}\\
&& \bolda'_2 \equiv s_1 s^*_4 + s_2 s^*_3,           \label{uhf-rfs0,2-2}\\
&& \zeta_{0+2}(X) \equiv \sum_{i=1}^4 s_i X s_i^*, \quad X\in\calO_{2,2},
                                                 \label{uhf-rfs0,2-3}\\[-10pt]
&& \varphi_{0+2}(X) \equiv \rho_{2,2}(X) 
   = \sum_{i=1}^4(-1)^{i-1}s_i X s_i^*, \quad X\in\calO_{2,2}, 
                                                 \label{uhf-rfs0,2-4}\\[-18pt]
&&\nonumber
\end{eqnarray}
which satisfy
\begin{alignat}{3}
&\mbox{$\phantom{\mbox{ii}}$i) seed condition:}&\ \ &
   \{\bolda'_i, \, \bolda'_ j \} = 0, \ \ 
   \{\bolda'_i, \, \bolda^{\prime\,\ast}_j\} = -\delta_{i,j}I, \ \ 
    i,\,j=1,\,2, \\
&\mbox{$\phantom{\mbox{i}}$ii) recursive condition:}&\ \ &
 \{\bolda'_i, \, \zeta_{0+2}(X) \}\!=\!0, \  
   \zeta_{0+2}(X)^*\!=\!\zeta_{0+2}(X^*),  \ X \in \calO_{2,2}, \\
&\mbox{iii) normalization condition:}&\ \ &
\zeta_{0+2}(X)\zeta_{0+2}(Y)=\varphi_{0+2}(XY), \quad X,\,Y\in\calO_{2,2}.
\end{alignat}
Then, an embedding $\varPhi_{R_{0+2}}$ of NCAR into $\calO_{2,2}$
associated with $SR_{0+2}$ is given by
\begin{equation}
\begin{array}{c}
\varPhi_{SR_{0+2}} : \mbox{NCAR} \hookrightarrow \calO_{2,2}, \\[5pt]
\varPhi_{SR_{0+2}}(a'_{2(n-1)+i}) 
   \equiv \zeta'{}^{n-1}(\bolda'_i), \quad i=1,\,2, \ \ n=1,\,2,\,\ldots.
\end{array}                                          \label{uhf-rfs0,2-5}
\end{equation}
Furthermore, we can show that $\varPhi_{SR_{0+2}}(\mbox{NCAR})$ 
is identical with $\calO_{2,2}^{U(1)}$ in the same way as 
$\varPhi_{SR_{1+1}}(\mbox{ICAR})$. 
This fact means that there exists a $\ast$-isomorphism between ICAR and 
NCAR in which metric structures are completely different from each other.
Indeed, using \eqref{PCR1} and \eqref{PCR2} with 
$\eta_{ij}=(-1)^{i-1}\delta_{ij}$, we obtain the relation between 
$(\bolda_1,\,\bolda_2)$ in $SR_{1+1}$ and
$(\bolda'_1, \, \bolda'_2)$ in $SR_{0+2}$ as follows
\begin{eqnarray}
\noalign{\vskip-2pt}
&& \bolda'_1 
   = \bolda_1  -  ( \bolda_1^*+ \bolda_1 ) \bolda_2^* \bolda_2, \\
&& \bolda'_2 
   = (\bolda_1^* - \bolda_1 ) \bolda_2\,; \\
\noalign{\eject}
&& \bolda_1 
   =  \bolda'_1  +  ( \bolda'_1{}^*+ \bolda'_1 ) \bolda'_2{}^* \bolda'_2, \\
&& \bolda_2 
   = (\bolda'_1{}^* - \bolda'_1 ) \bolda'_2. 
\end{eqnarray}
Then, we obtain the following one-to-one correspondence of generators 
between ICAR and NCAR:
\begin{eqnarray}
\noalign{\vskip-5pt}
a'_{2n-1} 
&\Leftrightarrow& 
  \exp(i\pi\sum_{k=1}^{n-1} a_{2k}^*\, a_{2k})
  \Big( a_{2n-1}  -  (a_{2n-1}^*+ a_{2n-1}) a_{2n}^*\, a_{2n}\Big),  \\[-2pt]
a'_{2n} 
&\Leftrightarrow& 
   \exp(i\pi\sum_{k=1}^{n-1} a_{2k-1}^*\, a_{2k-1})
   \Big( a_{2n-1}^* -  a_{2n-1} \Big) a_{2n}\,;  \\[-2pt]
a_{2n-1} 
&\Leftrightarrow&  
  \exp(i\pi\sum_{k=1}^{n-1} a^{\prime\,\ast}_{2k} \, a'_{2k})
  \Big(  a'_{2n-1}  
   +  ( a^{\prime\,\ast}_{2n-1} 
        + a'_{2n-1}) a^{\prime\,\ast}_{2n}\, a'_{2n}\Big), \\ [-2pt]
a_{2n} 
&\Leftrightarrow&  
  \exp(i\pi\sum_{k=1}^{n-1} a^{\prime\,\ast}_{2k-1} \, a'_{2k-1})
  \Big( a^{\prime\,\ast}_{2n-1} - a'_{2n-1} \Big) a'_{2n},   
\end{eqnarray}
where $\exp(i\pi a_n^* a_n)=I-(-1)^n2 a_n^*a_n$ and 
$\exp(i\pi a_n^{\prime\,\ast}a'_n)=I+2a_n^{\prime\,\ast}a'_n$ being 
the Klein operators anticommuting with $a_n$ and $a'_n$, respectively.
Under this nonlinear transformation, the vacuum of the Fock representation 
is kept invariant, but neither the particle number nor the metric structure 
is preserved.
Therefore, the difference between ICAR and NCAR is just by the choice of
generators corresponding to unitarily inequivalent representations.
Discovery of this kind of nonlinear transformation of the fermion algebra is
greatly indebted to the expressions of fermion generators in terms of the
pseudo Cuntz algebra.\cite{AK3}
The description of fermion algebras in the (pseudo) Cuntz algebra seems to 
play quite an important role in the study of fermion systems.
\vskip25pt
%
%
%
\app{Embeddings among Pseudo Cuntz Algebras}
Embeddings and endomorphisms considered in the ordinary Cuntz algebra 
may be easily generalized to the pseudo Cuntz algebra. 
In the similar way that an arbitrary Cuntz algebra $\calO_d$ is embedded 
into $\calO_2$ as its $\ast$-subalgebra,\cite{AK1,Cuntz} we can show that 
an arbitrary pseudo Cuntz algebra $\calO_{d,d'}$ is embedded into 
$\calO_{1,1}$.  
As a special case, $\calO_2$ is embedded into $\calO_{1,1}$.\footnote{In
precise, we consider only embeddings of the dense subalgebra of $\calO_2$ 
into $\calO_{1,1}$.}
\par
First, let us note the existence of the following operator $J$ in 
$\calO_{1,1}$: 
\begin{eqnarray}
&& J\equiv s_{2;\,1}-s_{1;\,2}, \\
&& J^* = -J, \quad J^* J = J J^* = - J^2 = - I.  
\end{eqnarray}
Then, it is straightforward to see that an embedding $\varPsi_2$ of 
$\calO_2$ into $\calO_{1,1}$ is given by
\begin{eqnarray}
&&
\begin{array}{c}
\varPsi_2 : \calO_2 \hookrightarrow \calO_{1,1}, \\[5pt]
\varPsi_2(S_1) \equiv s_1, \quad \varPsi_2(S_2) \equiv s_2\, J,
\end{array}
\end{eqnarray}
where $S_1$ and $S_2$ denote the generators of $\calO_2$.
Likewise, an embedding $\varPsi_d$ of $\calO_d$ into $\calO_{1,1}$ is given by 
\begin{eqnarray} 
&&
\begin{array}{c}
\varPsi_d: \calO_d \hookrightarrow \calO_{1,1}, \\[5pt]
\varPsi_d(S_i)
  = \begin{cases}
     s_2^{i-1}\,s_1\,J^{i-1} & \mbox{for $1\leqq i \leqq d-1$}, \\[5pt]
     s_2^{d-1}\,J^{d-1}      & \mbox{for $i=d$}. 
    \end{cases}
\end{array}
\end{eqnarray}
Now, it is easy to see that an embedding $\varPsi_{d,d'}$ of $\calO_{d,d'}$
with $d'\geqq1$ into $\calO_{1,\,1}$ is given by
\begin{eqnarray} 
&&
\begin{array}{c}
\varPsi_{d,d'}: \calO_{d,d'} \hookrightarrow \calO_{1,\,1}, \\[5pt]
\varPsi_{d,d'}(S_i) 
 = \begin{cases}
    s_2^{i-1}\,s_1\,J^{i-1} & \mbox{for $1\leqq i \leqq d$}, \\[5pt]
    s_2^{i-1}\,s_1\,J^{i}   & \mbox{for $d+1\leqq i \leqq d+d'-1$}, \\[5pt]
    s_2^{d+d'-1}\,J^{d+d'}  & \mbox{for $i=d+d'$}, 
   \end{cases}
\end{array}
\end{eqnarray}
with $S_i$'s being the generators of $\calO_{d,d'}$.
\par
As for $\calO_{2^p,2^p}$ with $p\geqq1$, there is an embedding into 
$\calO_{1,1}$ in which each generator of $\calO_{2^p,2^p}$ is mapped 
to an element in $\calO_{1,1}$ homogeneously in $s_i$ without using $J$ 
as follows:
\begin{eqnarray}
&&
\begin{array}{c}
\hat\varPsi_{2^p,d^p}: \calO_{2^p,2^p} \hookrightarrow \calO_{1,1}, \\[5pt]
\hat\varPsi_{2^p,2^p}(S_i)=s_{i_1}s_{i_2}\cdots s_{i_{p+1}}, \quad 
   i=1,\,\ldots,2^{p+1}, \ i_k=1,\,2,\  (k=1,\,\ldots,\,p+1)
\end{array}\quad
\end{eqnarray}
with an appropriate one-to-one correspondence between the indices $i$ 
and $(i_1,\,i_2,\,\ldots,\,i_{p+1})$.
\par
There is another type of embedding $\tilde\varPsi_2$ of  $\calO_2$ into 
$\calO_{1,\,1}$ as follows:
\begin{eqnarray}
&&
\begin{array}{c}
\tilde\varPsi_2: \calO_2\hookrightarrow\calO_{1,\,1},         \\[5pt]
\tilde\varPsi_2(S_1) \equiv \rho(s_1), \quad
\tilde\varPsi_2(S_2) \equiv \xi(s_2),                         \\[5pt]
 \begin{array}{l}
  \rho(X) 
    \equiv s_1\,X\,s_1{}^*-s_2\,X\,s_2{}^*,                   \\[5pt] 
  \xi(X) \equiv s_2\,X\,s_1{}^* - s_1\,X\,s_2{}^*,            \\[5pt]    
 \end{array}
\end{array}
\end{eqnarray}
where $\rho$ is nothing but the canonical endomorphism of 
$\calO_{1,1}$, and the mapping $\xi$ satisfies the following properties
\begin{eqnarray}
&& \xi(X)^* = - \xi(X^*),                            \label{xi1}\\
&& \xi(X)\xi(Y)=\rho(XY),                            \label{xi2}\\
&& \xi(X)\rho(Y)=\rho(X)\xi(Y)=\xi(XY).              \label{xi3}
\end{eqnarray}
It is also straightforward to generalize $\tilde\varPsi_2$ to the
corresponding embedding of $\calO_{d,d'}$ into $\calO_{1,1}$.
\vskip20pt
%
%
%

\end{document}